\documentclass[11pt,a4paper]{article}
\pdfoutput=1
\usepackage{jcappub}
\usepackage{graphicx}
\usepackage{subfig}

\def\apj{{\it Astrophys.~J.}}
\def\apjs{{\it Astrophys.~J. Suppl.}}

\def\apjl{{\it Astrophys.~J.~Lett.}}
\def\prd{{\it Phys.~Rev.~D}}
\def\prl{{\it Phys.~Rev.~Lett.}}
\def\plb{{\it Phys.~Letts.~B}}
\def\mnras{{\it Mon.~Not. Roy.~Astr.~Soc.}}
\def\ijmpd{{\it Int.~J.~ Mod. Phys. D}}

\def\AnA{{\it Astron. Astrophys.}}

\def\ARAnA{{\it Ann. Rev. Astron. Astrophys.}}

\title{Cosmokinetics: A joint analysis of Standard
  Candles, Rulers and Cosmic Clocks}
\author[a]{Remya Nair,}
\author[a]{Sanjay Jhingan}
\author[b]{and Deepak Jain}

\affiliation[a]{Centre for Theoretical Physics,\\ Jamia Millia
Islamia, New Delhi 110025, India}

\affiliation[b]{Deen Dayal Upadhyaya College, \\
University of Delhi, New Delhi 110015, India}

\emailAdd{remya$_{-}$phy@yahoo.com} \emailAdd{sjhingan@jmi.ac.in}
\emailAdd{djain@ddu.du.ac.in}

\abstract{
We study the accelerated expansion of the Universe by using
  the kinematic approach. In this context, we parameterize the
  deceleration parameter, $q(z)$, in a model independent way. Assuming
  three simple parameterizations we reconstruct $q(z)$. We do the joint
  analysis with combination of latest cosmological data consisting of
  standard candles (Supernovae Union2 sample), standard ruler (CMB/BAO),
  cosmic clocks (age of passively evolving galaxies) and Hubble
  ($H(z)$) data. Our results support the accelerated expansion of the
  Universe.}

\keywords{Cosmic acceleration, CMB/BAO, Lookback time, Hubble
parameter data, Supernovae}

\begin{document}

\maketitle

\section{Introduction}
Mapping the cosmic expansion history with accelerated expansion is
one of the major challenges faced by modern cosmology. Various
observations, directly (Type Ia Supernovae, SNe Ia) and indirectly
(CMB, BAO, lookback time) provide evidence for the recently observed
accelerated expansion of the Universe \cite{Riess, Perlmutter,
Astier, tegmark, eisen}. The phenomenon of acceleration is usually
attributed to some sort of dark energy  \cite{fri}. The effect of
dark energy is to change the sign of "deceleration parameter"
$q(z)$. Whether estimates of $q(z)$ point towards accelerated or
decelerated expansion, strongly depends on quality and quantity of
the observational data at various redshifts. Therefore, one of the
simple ways to understand the kinematics of the Universe is by
phenomenologically parameterizing $q(z)$.

The phenomenological approach is advantageous since it does not rely
on model specific assumptions like the composition of the Universe.
It is assumed that our Universe is homogeneous and isotropic at
large scales and is described by a metric theory of gravity. This
type of approach is explained in literature in a variety of ways
\cite{shap, blandford, turner, dodelson, german}.

Following the same line of thought we reconstruct the deceleration
parameter using the Supernovae Union2 data set, CMB/BAO, Hubble
parameter data (H(z)),  and lookback time (LBT). Since the nature of
the driving force of the Universe is still a mystery, therefore the
choice of parameterization of $q(z)$ is arbitrary. We use two
two-parameter models of $q(z)$ and and one one-parameter model to
reconstruct the deceleration parameter. Riess et al. \cite{re}
showed using Gold (SNe Ia) data set that $q(0) <0$ at $99\%$ confidence level,
and there was a transition from recent acceleration to past
deceleration. In this context the phenomenological approach to
parameterize the $q(z)$ is an easy way to find out $q(0)$, and the
transition redshift, $z_t$, where expansion switches from being
decelerated to accelerated.

In addition to the data sets used in previous works, we have used
the age of slowly evolving passive galaxies (lookback time) in our
analysis. The addition of the LBT is significant as it is
complementary to the observations used in previous works and may
help us to obtain more realistic constraints on the expansion
history.

The paper is organized as follows. In section II, we discuss the
$q(z)$ parameterizations, data and methodology. The results and
discussion are explained in section III.

\section{Model, Data and Methodology}
With the assumption that Universe is homogeneous and isotropic, and
spatial flatness which is motivated by inflation and WMAP
measurements \cite{komatsu09}, FRW metric describes the background
geometry
\[
ds^2 = -dt^2 + a(t)^2\left[dr^2+r^2 d\Omega^2 \right]\;.
\]
Here $a(t)$ is the scale factor. The cosmic scale factor is related
to the redshift of free streaming photons in the usual way: $a(t) =
1/(1+z)$ . The expansion and deceleration rates can be defined as
\begin{eqnarray}
H(z)&\equiv& \frac{\dot a}{a}, \\
q(z)&\equiv&-\frac{a\ddot{a}}{{\dot{a}}^2}=\frac{d}{dt}H^{-1}-1 \;,
\end{eqnarray}
where $H$ is the Hubble parameter. We can write the Hubble parameter
as
\begin{equation}
H=H_{0} \exp{\left[\int^{z}_{0}\frac{1+q(z')}{1+z'} dz'\right]}\;.
\end{equation}
Here $H_{0}$ is the present value of the Hubble parameter.

We parameterize the redshift dependence of $q(z)$ in the following
way,
\begin{align*}
q_{I}(z) &= q_{0}+q_{1}z \;,\\
q_{II}(z) &= q_{2}+q_{3}\frac{z}{1+z} \;, \\
q_{III}(z) &= \frac{1}{2} + \frac{q_{4}}{(1+z)^2} \;.
\end{align*}
The first parameterizations is a linear Taylor series expansion of
the deceleration parameter around $z=0$, where $q_{0}$ and $q_{1}$
are its present value and its first derivative, respectively. The
second parameterization has an advantage that it converges at high
redshift, as expected. The last parameterization is such that it
converges to the value $\frac{1}{2}$ at high redshift, which is the
value of the deceleration parameter at matter dominated epoch.

In general the dark energy density parameter can be expressed as a
function of redshift as
\begin{equation}
\Omega_{X}(z)=\Omega_{X0}\exp{\left[3\int^{z}_{0}\frac{1+w(z')}{1+z'}dz'\right]}\,
\end{equation}
where $w(z)$ is the equation of state and is related to $H(z)$ by
\begin{equation}
w(z)=\frac{\frac{2}{3} (1+z)\frac{dlnH}{dz}-1}{1-(\frac{H_0}{H})^2 \
\Omega_{m0} (1+z)^3 }.
\end{equation}
Substituting for $H(z)$ we can express $w(z)$ in terms of the
deceleration parameter q(z) as\cite{tarundeep}
\begin{equation}
w(z)=\frac{\frac{2}{3}
(1+q(z))-1}{1-\exp{\left[-2\int^{z}_{0}\frac{1+q(z')}{1+z'}
dz'\right]}\ \Omega_{m0} (1+z)^3}.
\end{equation}

\subsection{Lookback time}
Most of the observations used for constraining cosmological
parameters like SNe Ia, angular diameter distances etc. are distance
based measurements. The LBT on the other hand is based on ages of
distant galaxies.

The LBT to an object at redshift $z$ is defined as the difference
between the present age of the Universe and its age at redshift $z$
and can be calculated as
\begin{equation}
t_{L}(z,p)=H_{0}^{-1} \int_{0}^{z}\frac{dz^{'}}{(1+z^{'}) {\cal
H}(p)} \;,
\end{equation}
where ${\cal H}(p)$ is the dimensionless Hubble parameter ${\cal
H}(p) \equiv  H(p)/H_{0}$ and {\cal p} are the parameters of the
model. The observed lookback time $t_{L}^{obs}$ to an object at
redshift $z_{i}$ is defined as
\begin{equation}
t_{L}^{obs}(z_{i},t_{inc},t_{0}^{obs})=t_{0}^{obs}-t(z_{i})-t_{inc}\;,
\end{equation}
where $t_{0}^{obs}$ is the observed age of the Universe. $t(z_{i})$
is the age of the object defined as the difference between the age
of the Universe at redshift $z_{i}$ and the age of the Universe when
the object was born at redshift $z_{f}$.
\begin{equation}
t(z_{i},p) = H_{0}^{-1} \left[
  \int_{z_{i}}^{\infty}\frac{dz^{'}}{(1+z^{'}){\cal H}(p)}
  -\int_{z_{f}}^{\infty}\frac{dz^{'}}{(1+z^{'}) {\cal H}(p)} \right]
\end{equation}
or
\begin{equation}
t(z_{i},p) = H_{0}^{-1}
\int_{z_{i}}^{z_{f}}\frac{dz^{'}}{(1+z^{'}){\cal H}(p)}
\end{equation}
$t_{inc}=t_{0}^{obs}-t_{L}(z_{f})$ is the incubation time of the
object. Since we don't know the formation redshift of the objects in
our sample, $t_{inc}$ is treated as a nuisance parameter and we
marginalize over it.

{}From the definition of the LBT above, we can write
\begin{equation}
t_{L}(z,p)=H_{0}^{-1}
\int_{0}^{z}\frac{dz^{'}}{(1+z^{'})\exp\left(\int^{z_{'}}_{0}
\frac{1+q(u)}{1+u}~du\right)}
\end{equation}

Now, we can substitute for $q(z)$ and write the lookback time for
the three different parameterization. In the first case we get
\begin{equation}
t_{L}(z,p)=H_{0}^{-1} \int_{0}^{z} (1+z^{'})^{q_{1}-q_{0}-2}
\exp(-q_{1}z^{'})~dz^{'}
\end{equation}
For the second parameterization the lookback time can be written as
\begin{equation}
t_{L}(z,p)=H_{0}^{-1} q_{2}^{-q_{2}-q_{3}-1}
\left[\gamma(q_{2}+q_{3}+1,q_{3})-\gamma
\left(q_{2}+q_{3}+1,\frac{q_{3}}{1+z}\right)\right]
\end{equation}
where $\gamma$ is the incomplete gamma function. For the last case,
we get
\begin{equation}
t_{L}(z,p)=H_{0}^{-1} \exp \left(-\frac{q_{4}}{2}\right)
\int_{0}^{z} \frac{\exp\left(\frac{q_{4}}{2
(1+z^{'})^2}\right)}{(1+z^{'})^{5/2}}~dz^{'}
\end{equation}

To constrain the parameters we use the ages of 32 passively evolving
galaxies in the redshift interval $0.117 \leq z \leq1.845$
\cite{sim}. We assume a 12$\%$ one standard deviation uncertainty on
the age measurements. The value of the Hubble parameter $H_{0}$ is
fixed at $H_{0}=$ 74.2 km/sec/Mpc and the age of the Universe is
taken to be $t_{0}^{obs}=$ 13.75 $\pm$ 0.13 Gyr.

The likelihood function is defined as
\begin{equation}
{\cal L} \propto \exp\left(-\frac{\chi^{2}}{2}\right)
\end{equation}
where $\chi^2$ is given by
\begin{equation}
\chi^2(p,H_{0},t_{inc},t_{obs})=\sum_{i=1}^{32}
\frac{(t_{L}(z_{i},p,H_{0})-t_{L}^{obs}(z_{i},t_{inc},
t_{0}^{obs}))^{2}}{\sigma_{i}^{2}+\sigma_{t_{0}^{obs}}^{2}} +
\frac{(t_{0}(p,H_{0})-t_{0}^{obs})^{2}}{\sigma_{t_{0}^{obs}}^{2}}
\end{equation}
Here $\sigma_{i}$ is the uncertainty in the estimate of $t(z_{i})$,
$\sigma_{t_{0}^{obs}}$ is the uncertainty in the estimate of $t_{0}$
and  $t_{L}(z_{i},p,H_{0})$ and $t_{0}(p,H_{0})$ are the predicted
ages.

To marginalize over this nuisance parameter $t_{inc}$ we define a modified
log-likelihood function \cite{dan}
\begin{equation}
{\tilde \chi}^{2} =-2 \ln \int_{0}^{\infty}
\exp\left(-\frac{\chi^{2}}{2}\right)~dt_{inc}
\end{equation}
which reduces to
\begin{equation}
{\tilde \chi}^{2}_{_{LBT}}= A-\frac{B^{2}}{C}+D-2
\ln\left[\sqrt{\frac{\pi}{2C}}~ erfc \left(
\frac{B}{\sqrt{2C}}\right)\right]
\end{equation}
where
\begin{equation}
A = \sum_{i} \frac{\Delta^{2}}{\sigma_{T}^{2}},\quad B = \sum_{i}
\frac{\Delta}{\sigma_{T}^2},\quad C = \sum_{i}
\frac{1}{\sigma_{T}^{2}},\quad \Delta =
t_{L}(z_{i},p,H_{0})-[t_{0}^{obs}-t(z_{i})] .
\end{equation}
We minimize the chi-squared with respect to the model parameters in
the three cases to find the best fit values of the parameters.

\subsection{Hubble Parameter}
Measurements of the Hubble parameter as a function of redshift,
$H(z)$, can also be used to constrain the deceleration parameter.
Stern et al (2010) \cite{st}, gave 11 measurements of $H(z)$ in the
redshift range $0.1 \leq z \leq 1.75$. Further, Gaztanaga et al.
(2009) \cite{gz}, gave estimates of $H(z)$ determined from
line-of-sight BAO peak position observations. We use the combination
of these two samples to constrain our parameters. We have a total of
13 data points \cite{rat, zh, ma}.

For the three parameterizations the Hubble parameter can be written
in terms of the model parameter as follows:
\begin{align}
H(z) &= H_{0} ~ \exp(q_{1} z) ~ (1+z)^{1+q_{0}-q_{1}}\\
H(z) &= H_{0} ~ \exp\left(\frac{-q_{3} z}{1+z}\right) ~ (1+z)^{1+q_{2}+q_{3}}\\
H(z) &= H_{0} ~ \exp\left(\frac{q_{4} z}{1+z}\right)
\exp\left(\frac{-z}{2(1+z)}\right)~ (1+z)^{\frac{3}{2}}
\end{align}
We find the constraints on the model parameters by minimizing the
chi-squared function
\begin{equation}
\chi^{2}_{_{Hubble}}(p,H_{0}) =\sum_{i=1}^{13}
\frac{(H^{th}(z_{i},H_{0},p)-H^{obs}(z_{i}))^2}{\sigma^{2}_{H,i}}
\end{equation}

\subsection{Supernova Union2 data}
We use the Union2 compilation of 557 SNe Ia \cite{aman} for
comparing the observed luminosity distance (derived from the
distance modulus) with the theoretical luminosity distance. The
distance modulus and the luminosity distance are related as
\begin{equation}
\mu_{th} = m - M = 5 log_{10} \frac{d_{L}}{Mpc}+25
\end{equation}
Here $m$ and $M$ are the apparent and absolute magnitudes
respectively. The luminosity distance is related to the Hubble
parameter in a spatially flat Universe as
\begin{equation}
d_{L}(z)=(1+z) \int_{0}^{z} \frac{dz^{'}}{H(z^{'})}
\end{equation}
Substituting for the Hubble parameter in terms of the deceleration
parameter we get
\begin{align}
d_{L} &= \frac{(1+z)}{H_{0}} ~ \exp(q_{1}) ~~ q_{1}^{q_{0}-q_{1}} ~
\gamma(q_{1}-q_{0},q_{1},q_{1}(z+1))\\
d_{L} &= \frac{(1+z)}{H_{0}} ~ \exp(q_{3}) ~~ q_{3}^{-(q_{2}+q_{3})}
~
\gamma\left(q_{3}+q_{2},\frac{q_{3}}{1+z},q_{3}\right)\\
d_{L} &= \frac{(1+z)}{H_{0}} \int_{0}^{z} (1+z^{'})^{-3/2}
\exp\left(\frac{q_{4}}{2 (1+z^{'})^2}\right)
\exp\left(\frac{-q_{4}}{2}\right) ~dz^{'}
\end{align}
in the case of the three parameterizations. Here $\gamma$ is the
generalized incomplete gamma function. We find the constraints on
the model parameters by minimizing the chi-squared function
\begin{equation}
\chi_{_{Union2}}^2 (p,H_{0}) =\sum_{i=1}^{557}
\frac{(d_{L}^{th}(z_{i},H_{0},p)-d_{L}^{obs}(z_{i}))^2}{\sigma^{2}_{d_{L},i}}
\end{equation}

\subsection{CMB/BAO}
BAO refers to a length scale in the distribution of photons and
baryons by the propagation of sound waves in the plasma of the early
Universe and they can be treated as cosmological standard rulers.
\cite{eisen, Cooray}. The distilled parameter $d_{z}$ \cite{wz1,wz2}
 is defined as $d_{z}\equiv \frac{r_{z}(z_{d})}{D_{V}(z)}
$, where $r_{s}(z_{d})$ is the comoving sound horizon size at the
baryon drag epoch and $D_{V}$ is the `dilation scale' distance given
by
\begin{equation}
D_{V}(z)=\left((1+z)^2 D_{A}(z)^2 \frac{c
z}{H(z)}\right)^\frac{1}{3}
\end{equation}
where $D_{A}$ is the angular diameter distance given by
\begin{equation}
D_{A}(z)=\frac{1}{(1+z)} \int_{0}^{z} \frac{dz^{'}}{H(z^{'})}
\end{equation}
We can further use the measurement of the acoustic scale $l_{A}$
provided by CMB to define a ratio $R_{z}$ as
\begin{equation}
\frac{1}{R_{z}}\equiv \frac{l_{A} d_{z}}{\pi} =
(1+z_{*})\frac{D_{A}(z_{*})}{r_{s}(z_{*})}
\frac{r_{s}(z_{d})}{D_{V}(z)}
\end{equation}
The value of the ratio between the sound horizon at last scattering
and at the baryon drag epoch is nearly 1.044 \cite{wz2}. So we can write
\begin{equation}
\frac{1}{R_{z}}\equiv \frac{l_{A} d_{z}}{\pi} \approx
(1+z_{*})~1.044~\frac{D_{A}(z_{*})}{D_{V}(z)}
\end{equation}
where $z_*$ is the redshift of recombination, $r_{s}(z_*)$ is the size of the sound horizon at last scattering and $D_A(z_*)$
is the physical angular diameter distance at the decoupling surface.
 We use six data points from SDSS LRG, 6dFGS and WiggleZ surveys. Three data
 points at high redshift ($z=0.44, 0.6, 0.73$) are from WiggelZ survey \cite{wz1}.
Two data points at redshift ($z=0.2, 0.35$) are from SDSS LRG survey (Percival
et al. \cite{percival}). The ($z= 0.106$) 6dFGS data point at low redshift is
taken from Beutler et al. \cite{beutler}. We derive the values of
${1}/{R_{z}}$ at 6 redshift points and find the corresponding
errors by using the data and the correlation coefficients provided
by Blake et al. \cite{wz1}. We use $l_{A}$=302.09 $\pm$0.76
\cite{komatsu11}.

The chi-squared function is given by
\begin{equation}
\chi^{2}_{_{CMB/BAO}}=(f^{obs}-f^{th})^T C^{-1}(f^{obs}-f^{th})
\end{equation}
where $f\equiv\frac{1}{R_{z}}$ and C is the covariance matrix
evaluated using the correlation coefficients.
The combined $\chi^2$ is given by
\begin{equation}
\chi^2_{combined} = \chi^2_{_{Union2}}+\chi^2_{_{Hubble}}+{\tilde
\chi}^2_{_{LBT}}+\chi^2_{_{CMB/BAO}}
\end{equation}

\subsection{Estimating parameter errors}

The method for estimating errors on parameter values is given below.
The 1$\sigma$ marginalized uncertainties on the two parameters and
their covariance can be calculated by using the following formulae
\cite{sivia}:
\begin{equation}
\sigma_{q_{0}}^2 =\frac{b}{c^{2}-ab}, \quad \sigma_{q_{1}}^2
={\frac{a}{c^{2}-ab}}, \quad \sigma_{q_{0} q_{1}}^2
={\frac{c}{c^{2}-ab}}\;,
\end{equation}
where
\[
a = \left.\frac{\partial^{2} L}{\partial q_{0}^{2}}
\right|_{q_{0}^{f},q_{1}^{f}},\quad b = \left.\frac{\partial^{2}
L}{\partial q_{1}^{2}} \right| _{q_{0}^{f},q_{1}^{f}}, \quad c =
\left.\frac{\partial^{2} L}{\partial q_{0} \partial q_{1}} \right|
_{q_{0}^{f},q_{1}^{f}} \;.
\]
Here $L$ is the log-likelihood function, and $q_{0}^{f}$ and
$q_{1}^{f}$ are the best fit parameter values.

\section{Results and Discussion}

In this work we follow a model independent methodology  to
reconstruct the expansion history of the Universe. The significance
of this kinematic approach lies in its simplicity because no
dependence on the matter-energy contents of the Universe is assumed
and further this approach does not demand any specific theory of
gravity. This idea has been used in the past by many authors to
prove the transition of the Universe from deceleration to
acceleration phase.

In 2002 by assuming a piecewise constant acceleration model with two
distinct epochs, Turner and Riess showed that Universe will
accelerate today if the transition redshift is fixed between 0.4 and
0.6 \cite{turner}. In a seminal work by Reiss et al. (2004), it was
shown that Universe underwent transition from deceleration to
acceleration by assuming the linear parameterization of $q(z)$
\cite{re}. Further by using Gold SNe Ia data,  Shapiro and Turner
applied principal component analysis of $q(z)$ and found very strong
evidence ($5\sigma$) for the acceleration of the Universe in the
recent past \cite{shap}. Elgaory $\&$ Multamuki (2006)
\cite{elgaroy}, Gong $\&$ Wang (2006, 2007) \cite{gong}, Cunha $\&$
Lima (2008) \cite{cun}, and Guimares, Cunha $\&$ Lima (2009)
\cite{gui} again used SNe Ia data to map the kinematic expansion
history of the Universe.

Rapetti et al (2007) \cite{rapetti} developed new kinematical
technique to study the expansion history of the Universe. They used
the parameter space defined by the present value of deceleration
parameter and the jerk parameter, $j$ (dimensionless third
derivative of scale factor w.r.t. cosmic time). By using SNe Ia data
and X-ray gas mass fraction measurements they measured $ q_0 =
-0.82\pm 0.14$. Recently Lima, Holanda $\&$ Cunha (2009,2010)
\cite{lim} used the Sunyaev- Zeldovich effect and X-ray surface
brightness data to study the kinematical description of the
expansion of the Universe.

In this work we also followed the same line of thought and
reconstructed $q(z)$ by following purely kinematic approach.

\begin{itemize}

\item{\bf Parameterization I}: $ q_{I}(z) = q_0 + q_1 \,z$

The $1\sigma$, $2\sigma$ and $3\sigma$ confidence contours in the
$q_0 - q_1$ plane are shown for the combined data sets, both with
LBT and without LBT data in Fig \ref{fig11} and Fig \ref{fig12}
respectively. The contours obtained with joint analysis are very
tight as compared to the ones obtained with SNe and galaxy clusters.
The best fit values of the model parameters and the $z_t$ for
different data sets are displayed in the table \ref{firstpar} below.
The best fit values with 1$\sigma$ errors for
(CMB/BAO+Hubble+Union2+LBT) data set and for (CMB/BAO+Hubble+Union2)
data set are ($q_0 = -0.332 \pm 0.018$, $q_1=0.146 \pm 0.011$,
$z_t=2.27 \pm 0.12$) and ($q_0 = -0.371 \pm 0.023$, $q_1=0.154 \pm
0.011$, $z_t=2.4 \pm 0.22$) respectively. The value of $w(z)$ at
present with $1\sigma$ errors for (CMB/BAO+Hubble+Union2+LBT) data
set is $-0.792 \pm 0.017$.

\begin{table}[h]
  \caption{Best fit values for first parameterization}
      \centering
  \begin{tabular}{| c | c | c | c | c | c |}
    \hline
    {Data Set} &  $\chi^2/d.o.f$ & $q_0$ & $q_1$ & {$q(0)$} &{$z_{t}$} \\
    \hline
   CMB/BAO  & 0.538 &
    $
    -0.491$ & $
    0.202$ & -0.491  & 2.43 \\
   Union2  & 0.954 & $
    -0.524$ & $
    0.859$ & -0.524   & 0.61 \\
   Hubble  & 0.765 & $
    -0.347$ & $
    0.449$ &  -0.347 & 0.77\\
   LBT & 0.491 & $
    -0.202$ & $
    0.07$ & -0.202 & 2.88\\\hline
   CMB/BAO+Union2+Hubble & 0.971 & $
    -0.371$ & $
    0.154$ & -0.371 & 2.4
    \\ \hline
   CMB/BAO+Union2+ Hubble + LBT & 1.009 & $
    -0.332$ & $
    0.146$ & -0.332 & 2.27
    \\
    \hline
  \end{tabular}\label{firstpar}
\end{table}

Fig.\ref{fig31} and Fig.\ref{fig33} show the evolution of
deceleration parameter and equation of state respectively, with
redshift $z$. There is a transition from accelerated to decelerated
phase. The transition redshift in this model is $z_t = 2.4$ (with
SNe Ia + H(z) + CMB/BAO) and the $z_t$ is lower down to 2.2 when LBT
data is included in the joint analysis of SNe Ia, H(z) and CMB/BAO.
This $z_t$ is very high as compare to $\Lambda$CDM prediction ($z_t
= 0.66$). The joint analysis with LBT gives present value of
deceleration parameter as $ q_0 = -0.332$.

\begin{figure}
\centering  \subfloat[Part 1][Without lookback
time]{\includegraphics[bb=0 0 288
286,width=2.5in]{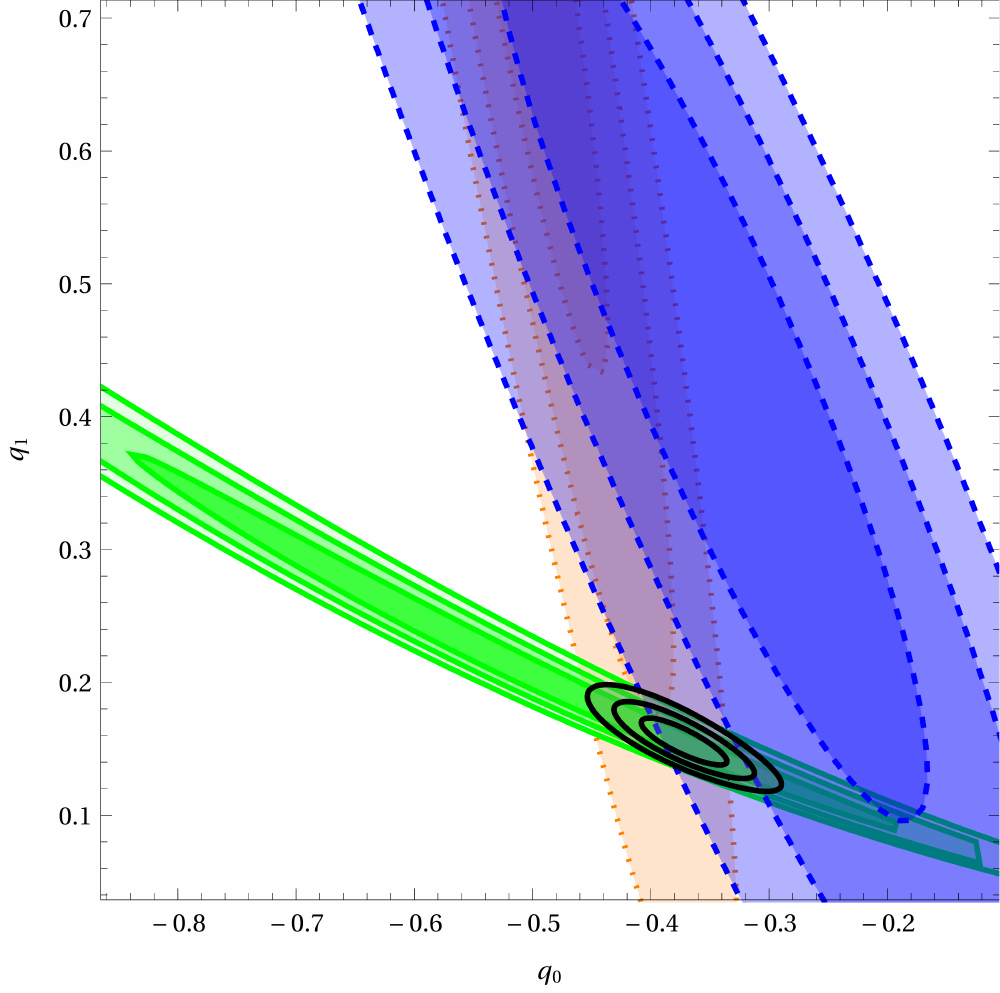}\label{fig12}} \subfloat[Part 1][With lookback
time]{\includegraphics[bb=0 0 288 286,width=2.5in]{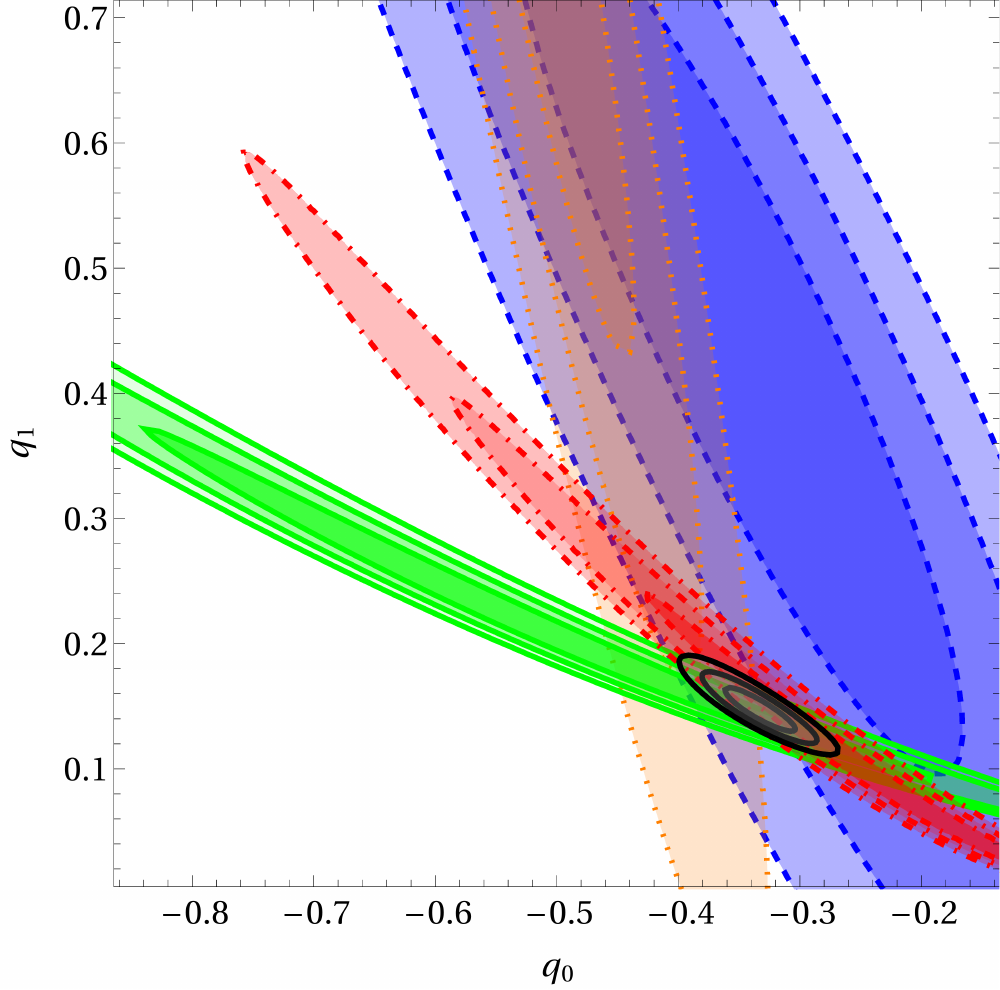}
\label{fig11}} \caption{The $1
\sigma, 2\sigma $ and $3\sigma$ contours in $q_{0} - q_{1}$ for
first parameterization. The gray contours are for the combined
chi-squared. Green (continuous line) contours correspond to CMB/BAO,
red (Dash-dot) curves are for lookback time, orange (Dotted) lines
are SNe Ia and blue (Dashed) curves are for Hubble data.}
\end{figure}

\item {\bf Parameterization II}: $ q_{II}(z) = q_2 + q_3(\frac{z}{1+z})$

Confidence regions ($1\sigma$, $2\sigma$ and $3\sigma$) in $q_2-
q_3$ parametric plane for SNe Ia, H(z), CMB/BAO and joint analysis
are shown in Fig \ref{fig22} and Fig. \ref{fig21}. The best fit
values with 1$\sigma$ errors for the (CMB/BAO + Hubble + Union2 +
LBT) data set and for (CMB/BAO + Hubble + Union2) are ($q_2 = -0.595
\pm 0.073$, $q_3=1.278 \pm 0.042$, $z_t=0.87 \pm 0.20$) and ($q_0 =
-0.526 \pm 0.028$ , $q_3=1.205 \pm 0.046$, $z_t=0.77 \pm 0.09$)
respectively. The value of $w(z)$ at present with $1\sigma$ errors
for (CMB/BAO+Hubble+Union2+LBT) data set is $-1.042 \pm 0.07$.

\begin{table}[h]
  \caption{Best fit values for second parameterization}
      \centering
  \begin{tabular}{| c | c | c | c | c | c |}
    \hline
    {Data Set} &  ${\chi}^2/d.o.f$ & ${q_2}$ & ${q_3}$ & {$q(0)$} & {$z_{t}$}\\
    \hline
   CMB/BAO  & 0.556 &
    $
    -0.775$ & $1.578$ & -0.775 & 0.96 \\
   Union2  & 0.956 & $
    -0.552$ & $
    1.335$  & -0.552 & 0.70\\
   Hubble  & 0.761 & $
    -0.411$ & $
    1.014$ & -0.411 & 0.68\\
   LBT & 0.495 & $ -0.325$ &
    $0.58$ & -0.325 & 1.27 \\ \hline
   CMB/BAO + Union2 + Hubble & 0.947 & $
    -0.526$ & $
    1.205$ & -0.526 & 0.77\\\hline
   CMB/BAO + Union2 + Hubble + LBT & 0.957 & $ -0.595$ &
    $1.278$ & -0.595 & 0.87\\\hline
  \end{tabular}\label{secpar}
\end{table}

Similarly the joint analysis is further done by the inclusion of LBT
for this parameterization, shown in Fig. \ref{fig21}. The joint
analysis with LBT shifts the best fit parameter values towards the
higher side. The constraints obtained on the parameter values by the
joint analysis are very tight as compared to the constraints obtain
from the SNe Ia data and the galaxy cluster data sets independently.

Fig.\ref{fig32} and Fig.\ref{fig34} display the variation of
deceleration parameter and equation of state respectively, w.r.t.
the redshift. The reconstruction of the $q(z)$ is done by the joint
analysis of SNe Ia + CMB/BAO + LBT + H(z) data sets. The central
line is drawn with the best fit values of the model parameters. The
transition redshift in this case is, $ z_t = 0.87$.

\begin{figure}
\centering
\subfloat[Part 2][Without lookback time]{\includegraphics[bb=0 0 288
286,width=2.5in]{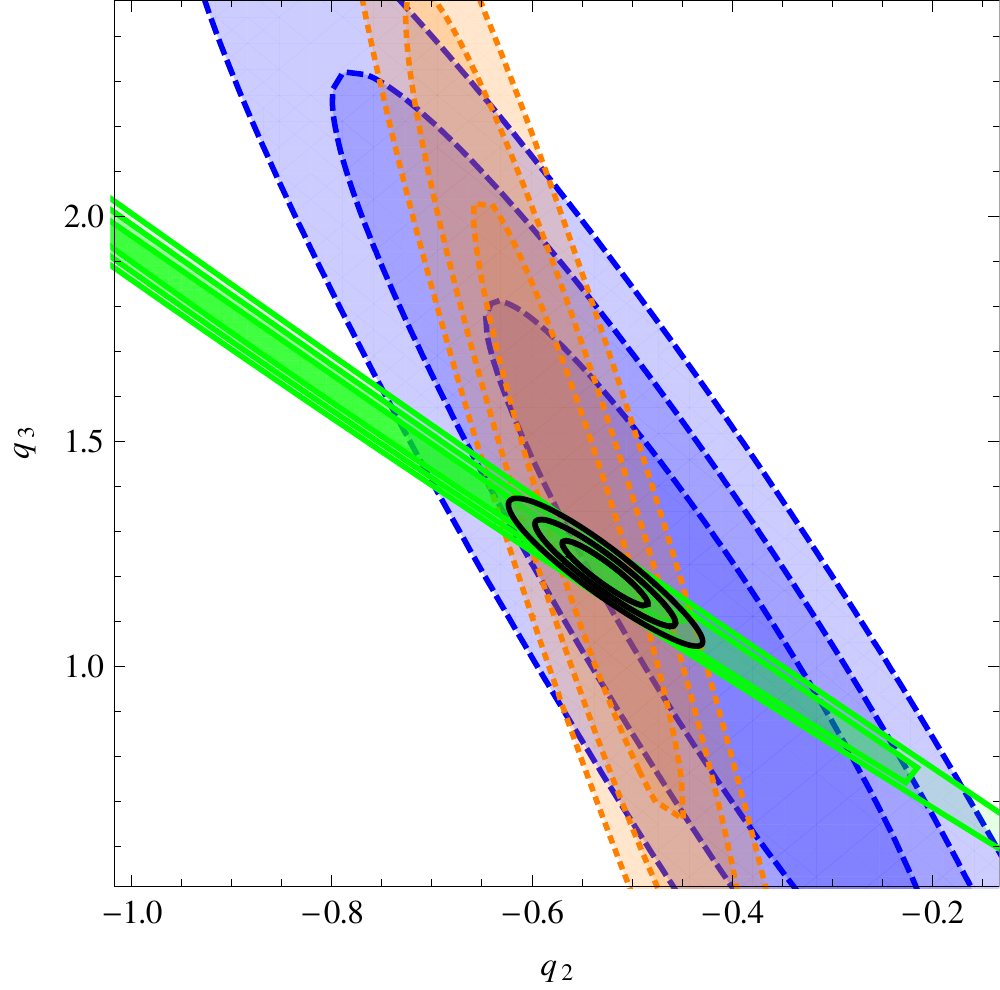}\label{fig22}} \subfloat[Part 2][With lookback
time]{\includegraphics[bb=0 0 288
286,width=2.5in]{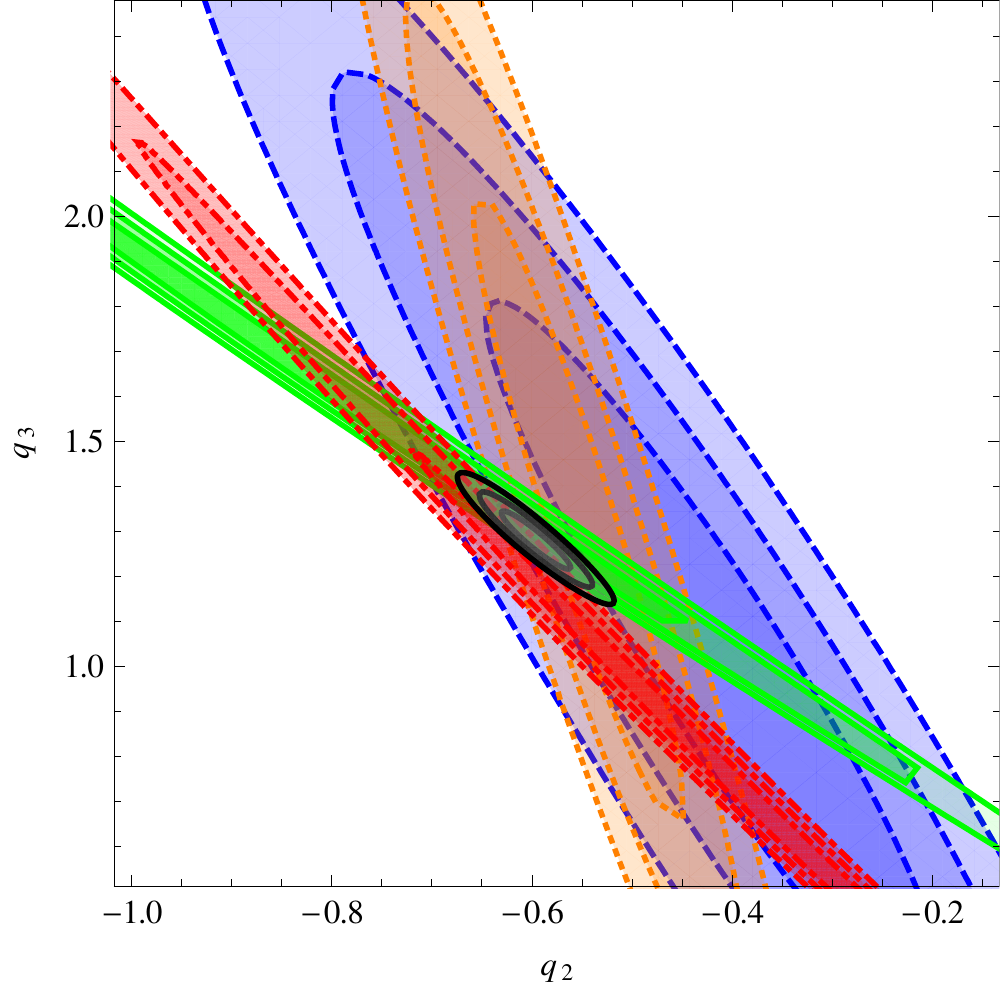}\label{fig21}} \caption{The $1
\sigma, 2\sigma$ and $3\sigma$ contours in $q_{2} - q_{3}$ plane
with all the data sets in case of second parameterization. The gray
contours are for the combined chi-squared. Green (continuous line)
contours correspond to CMB/BAO, red (Dash-dot) curves are for
lookback time, orange (Dotted) lines are SNe Ia and blue (Dashed)
curves are for Hubble data.}
\end{figure}

\begin{figure}[ht]
\centering \subfloat[Part 3][]{
\includegraphics[bb=0 0 288
286,width=3in]{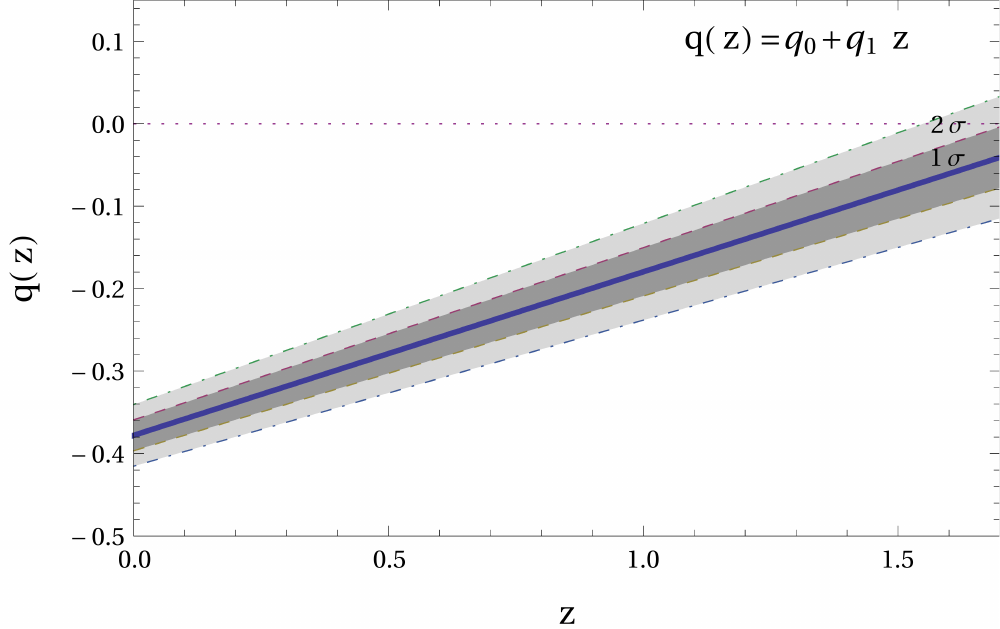}\label{fig31}} \subfloat[Part
3][]{\includegraphics[bb=0 0 288
286,width=3in]{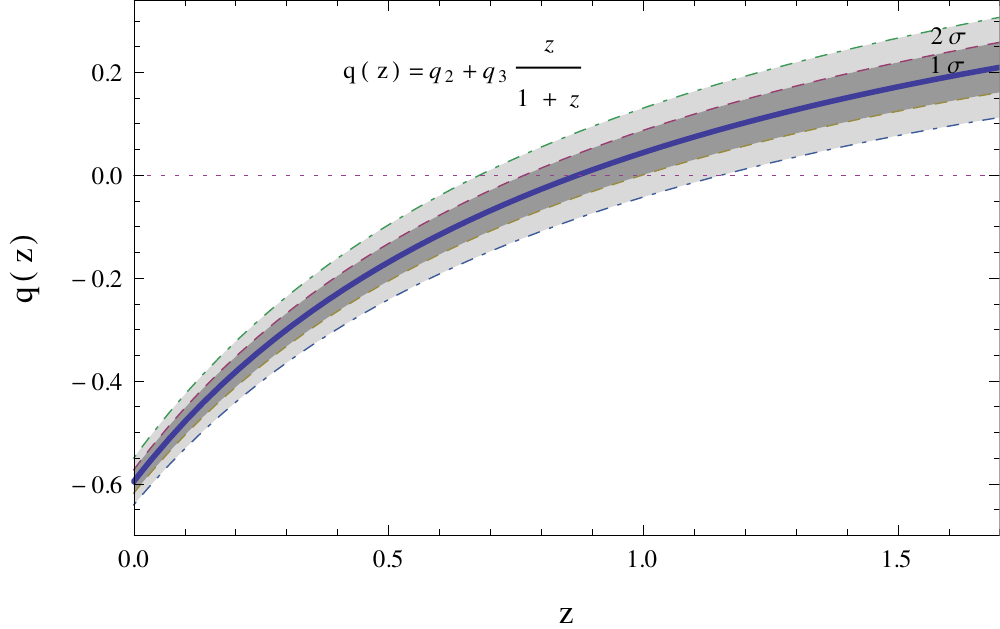}\label{fig32}} \\ \vspace{-1in}\centering
~~~~\subfloat[Part 3][]{
\includegraphics[bb=0 0 288
286,width=3.1in]{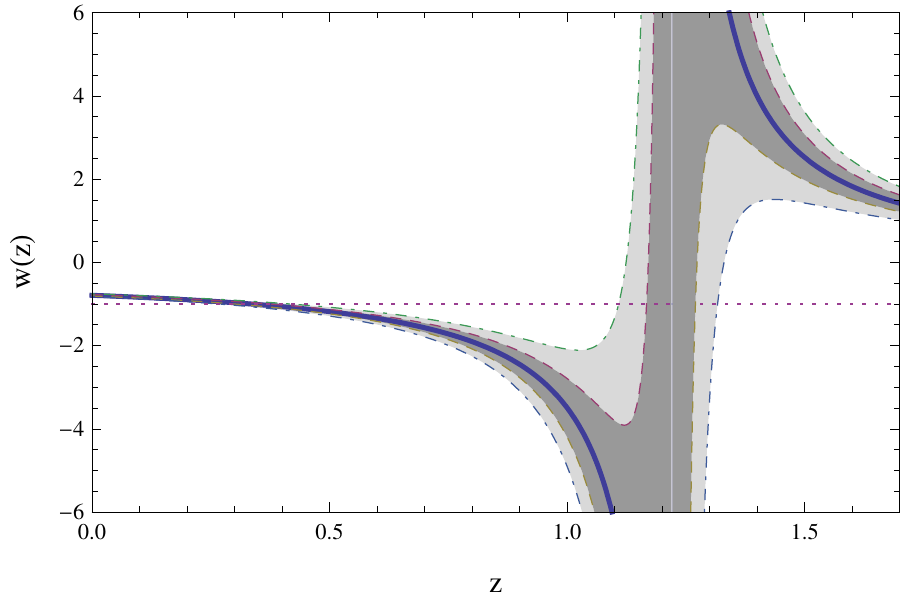}\label{fig33}} \subfloat[Part
3][]{\includegraphics[bb=0 0 288
286,width=3.1in]{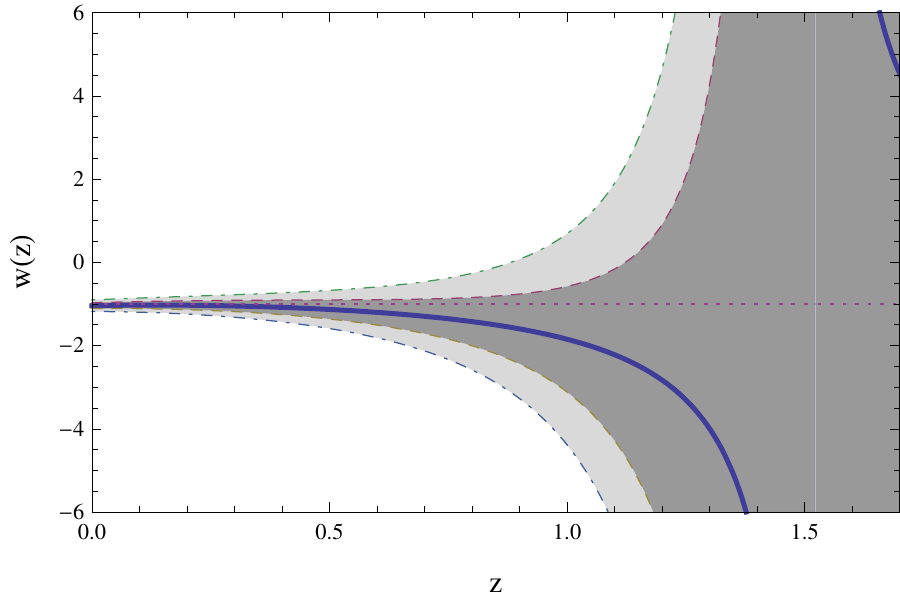}\label{fig34}} \caption{(a) and (b)
correspond to variation of q(z) vs z for combined chi-squared for
first  and second parameterization respectively with $1 \sigma$ and
$2 \sigma$ level. (c) and (d) correspond to variation of $w(z)$ vs z
for combined chi-squared for first and second parameterization
respectively with $1 \sigma$ and $2 \sigma$ level. For plotting (c)
and (d) the value of $\Omega_{m0}$ is taken to be 0.3.}
\end{figure}

\item {\bf Parameterization III}: $ q_{III}(z)= \frac{1}{2} +
\frac{q_4}{(1+z)^2}$
The best fit values with 1$\sigma$ errors for the
 (CMB/BAO + Hubble + Union2 + LBT) data set and for (CMB/BAO + Hubble + Union2)
 are ($q_4 = -1.296 \pm 0.024$, $z_t=0.61 \pm 0.01$) and ($q_4 = -1.162 \pm
 0.03$, $z_t=0.52 \pm 0.02$)
 respectively  and
hence the present value of deceleration parameter equal to  $ q_0 =
-0.796$ (See Figs \ref{fig41} and \ref{fig42}). The value of $w(z)$
at present with $1\sigma$ errors for (CMB/BAO + Hubble + Union2 +
LBT) data set is $-1.234 \pm 0.023$.

The best-fit evolution of deceleration parameter and equation of
state with redshift is shown in Fig \ref{fig43} and Fig \ref{fig44}
respectively. The $1\sigma$ error bar in this curve is very tight as
compare to the two parameter model of parameterization. The
transition redshift in this case is $ z_t = 0.61$ which is in
agreement with $\Lambda$ CDM model with in 1$\sigma$ level. The
results for single parameter parameterization for the rest of the
data sets are summarized in table \ref{thirdpar} below.

\begin{table}[h]
  \caption{Best fit values for third parameterization}
      \centering
  \begin{tabular}{| c | c | c | c | c |}
    \hline
    {Data Set} &  {$\chi^2/d.o.f$} & {$q_4$} & {$q(0)$} & {$z_{t}$} \\
    \hline
   CMB/BAO  & 0.589 &
    $
    -1.264$ & -0.764 & 0.58\\
   Union2  & 0.957 & $-1.14
     $ & -0.64 & 0.51\\
   Hubble  & 0.805 & $
    -1.115$ & -0.615 & 0.49\\
   LBT &  0.567 & $-1.47 $ & -0.97 & 0.71\\\hline
   CMB/BAO + Union2 + Hubble & 0.952 & $-1.162 $ & -0.66 & 0.52\\\hline
   CMB/BAO + Union2 + Hubble + LBT & 1.028 & $-1.296  $ & -0.796 &
   0.61 \\\hline
  \end{tabular}\label{thirdpar}
\end{table}

\begin{figure}[ht]
\centering \subfloat[Part 4][Without lookback
time]{\includegraphics[bb=0 0 288
286,width=2.7in]{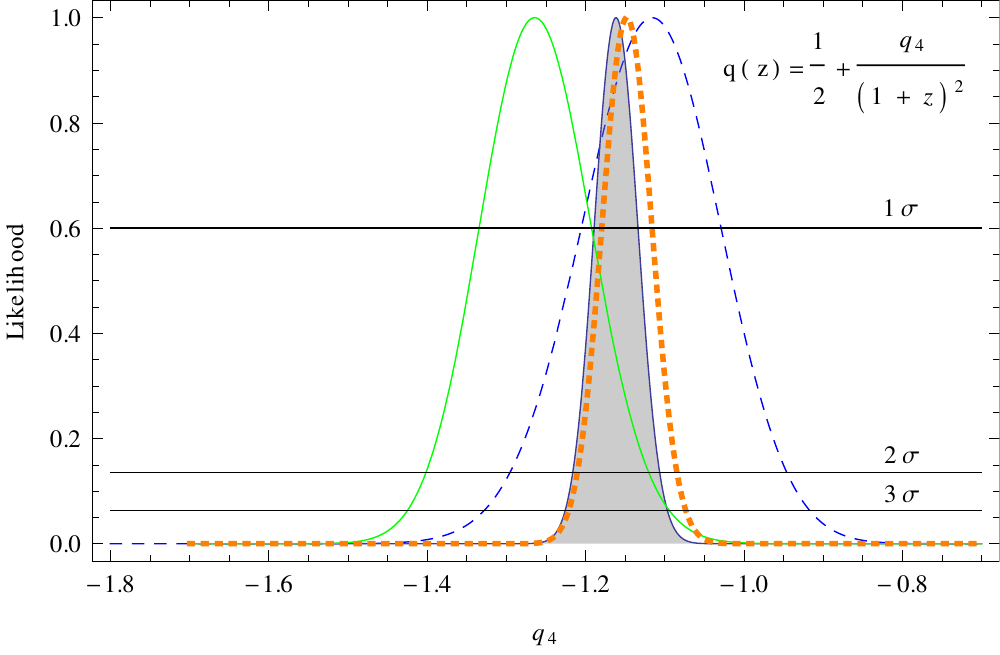}\label{fig42}} \subfloat[Part 4][With
lookback time]{\includegraphics[bb=0 0 288
286,width=2.7in]{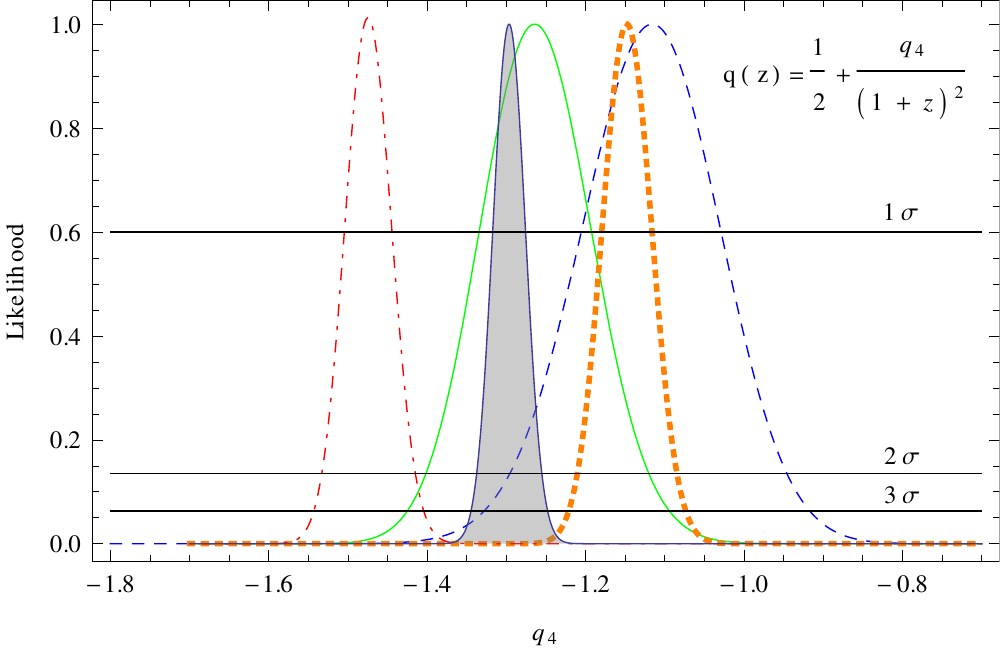}\label{fig41}}  \\ \vspace{-1.0 in}
~~\subfloat[Part 4][]{\includegraphics[bb=0 0 288
286,width=2.8in]{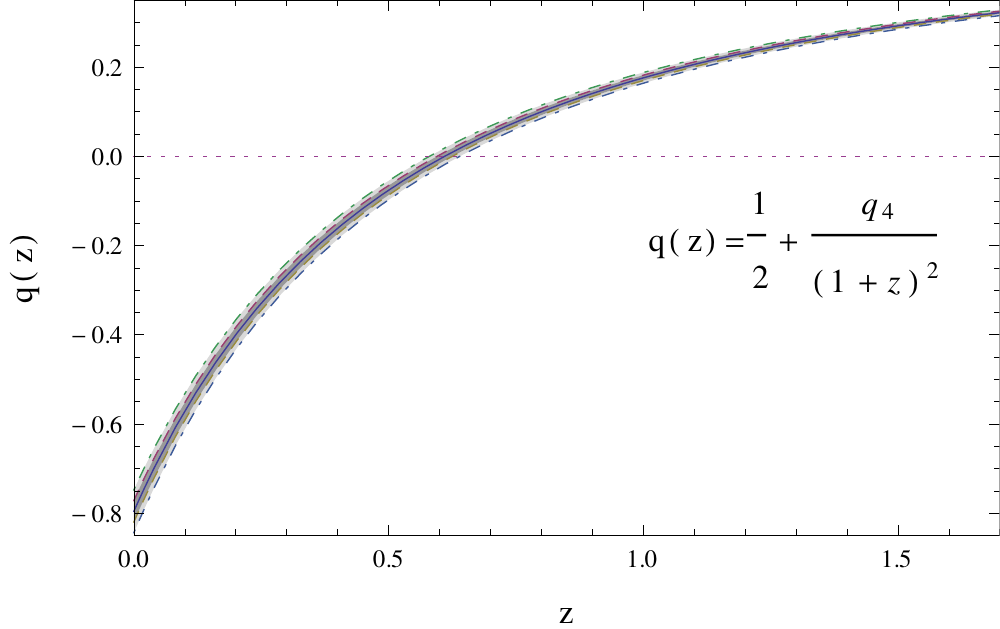}\label{fig43}} \subfloat[Part
4][]{\includegraphics[bb=0 0 288
286,width=3in]{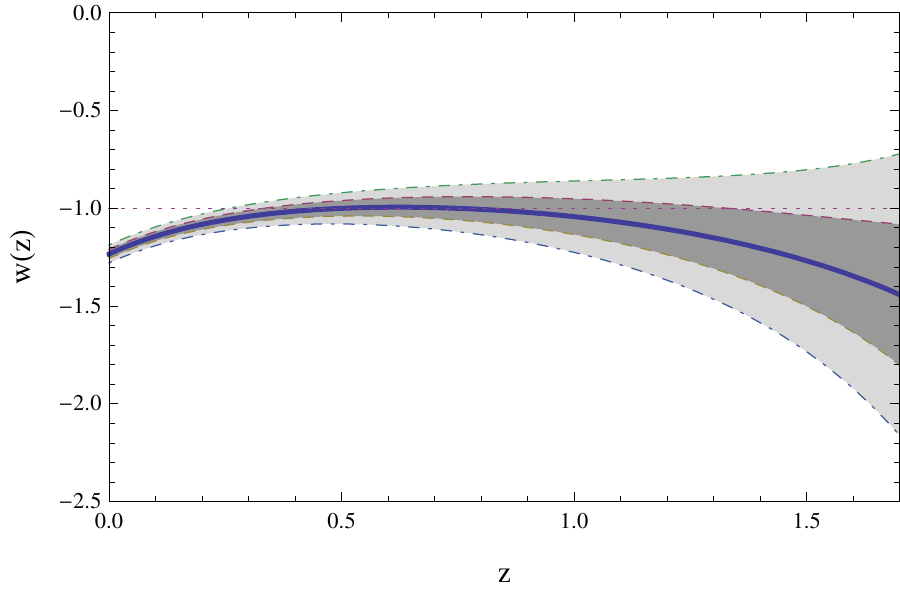}\label{fig44}} \caption{ (a) and (b),
Likelihood vs parameter plots with all the data sets in case of
third parameterization. The filled gray plot are for the combined
chi-squared. Green (continuous line) correspond to CMB/BAO, red
(Dash-dot) curves are for lookback time, orange (Dotted) lines are
SNe Ia and blue (Dashed) curves are for Hubble data. (c) q(z) vs z
for combined chi-squared for third parameterization with $1 \sigma$
and $2 \sigma$ level (d) $w(z)$ vs z for combined chi-squared for
third parameterization with $1 \sigma$ and $2 \sigma$ level. For
plotting (c) and (d) the value of $\Omega_{m0}$ is taken to be 0.3.}
\end{figure}

\end{itemize}

The summary of the results are the following:

\begin{itemize}
\item We have shown that inclusion of LBT in the joint analysis is
very important in the measurement of cosmological parameters. This
addition of age of passively evolving galaxies as a cosmic clock is
completely independent and competitive with standard candles and
rulers.

\item For all the three parameterizations considered here, the data
favors a transition from the deceleration phase to acceleration
phase of the Universe. Also, at present our Universe is
accelerating. This is also supported by the value of equation of
state of dark energy at the present epoch (see Figs 3(c), 3(d) and
4(d)).

\item The transition redshift $z_t$ strongly depends upon the form of
the parameterization of $q(z)$. The transition redshift from
decelerated to accelerated expansion ($q(z) = 0$) increases with the
inclusion of LBT in the analysis except for the parameterization I.
Similarly the present value of deceleration parameter become more
negative with the inclusion of LBT except for the parameterization
I. But the change in the value of $q_0$ is small.

\item The linear parameterization, $q_{I}$,  predicts the transition
redshift $z_t > 2$ which is not compatible with the  $\Lambda$CDM
model even at $3\sigma$ level. So this model is not reliable at all.

\end{itemize}

Using SNe Ia data only, Mortsell $\&$ Clarkson (2009) \cite{mort}
showed that the expansion of Universe is accelerating at low
redshifts even at $ > 12 \sigma$ level. In their recent work Xu et
al. \cite{xu} put bound on model parameters of $q(z)$ by using SNe
Ia, BAO and observational H(z) data. Further Lu et al.
(2011)\cite{lu} again tried to constrain the kinematic model by
using the SNe Ia and H(z) data only. They show the two parameterized
forms of $q(z)$ clearly deviated from the $\Lambda$CDM model.

It is natural to extend this work with addition of more observational
data set like gravitational lensing as a new standard ruler, GRB's as
a standard candles and age of globular clusters as a cosmic
chronometers in the present work.

\section*{Acknowledgement} We thank Tarun Deep Saini for useful
discussions and Florian  Beutler for suggestions. One of the author
(DJ) thanks Prof. A. Mukherjee and Prof. S. Mahajan for providing
the facilities to carry out the research. RN acknowledges support
under CSIR - JRF scheme (Govt.of India). Authors acknowledge the
financial support provided by Department of Science and Technology,
India under project No. SR/S2/HEP-002/2008.


\begin{thebibliography}{100}

\bibitem{Riess} Riess A. G. et al., (Supernova Search Team) {\it Observational evidence
from Supernovae for an accelerating Universe and a cosmological
constant}, \apj~
 {\bf 116} (1998) 1009,
[\texttt {arXiv:astro-ph/9805201}].

\bibitem{Perlmutter}  Perlmutter S. et al., (Supernova Cosmology Project)
{\it Measurement of Omega and Lambda from 42 high-redshift Supernovae},
\apj~ {\bf 517}  (1999)  565, [\texttt {arxiv:astro-ph/9812133}].

\bibitem{Astier}  Astier P. et al., {\it The Supernovae Legacy Survey: measurement of
$\Omega_m$, $\Omega_\lambda$ and $w$ from the first year data set}, \AnA ~ {\bf 447}
(2006) 31, [\texttt {arXiv:astro-ph/0510447}].

\bibitem{tegmark} Tegmark M. et al., (SDSS) {\it Cosmological parameter from SDSS \& WMAP},
\prd~{\bf 69}  (2004) 103501, [\texttt {arXiv:astro-ph/0310723}].

\bibitem{eisen} Eisenstein et al., {\it Cosmic Complementarity: $H_0$ and
$\Omega_m$ from Combining CMB Experiments and Redshift Surveys},
\apj~ {\bf 504} (1998) 57, [\texttt {arXiv:astro-ph/9805239}].

\bibitem{fri} Padmanabhan T., \textsl{Cosmological constant:
The weight of the vacuum, Phys. Rep.} {\bf 380} (2003) 235 [\texttt
{arXiv:hep-th/0212290}]; Copeland E. J., Sami M.\&  Tsujikawa S.,
\textsl{Dynamics of dark energy, \ijmpd}~{\bf 15} (2006) 1753
[\texttt {arXiv:hep-th/0603057}]; Sahni V. \& Starobinsky A.,
\textsl{Reconstructing Dark Energy, \ijmpd} {\bf 15} (2006) 2105
[\texttt {arXiv:astro-ph/0610026}] ; Frieman J. A., Turner M. \&
Huterer D., \textsl{Dark Energy and the Accelerating Universe,
\ARAnA}~ {\bf 46} (2008) 385 [\texttt {arXiv:0803.0982}]; Caldwell
R. R. \& Kamionkowski M., {\it The Physics of Cosmic Acceleration,
Ann. Rev. Nucl. Part. Sci.} {\bf 59} (2009) 397 [\texttt
{arXiv:0903.0866}] ; Li M. et al., {\it Dark Energy}, {\it Commun.
Theor. Phys} {\bf56} (2011) 525.

\bibitem{shap} Shapiro C. \& Turner M., {\it What Do We Really Know
About Cosmic Acceleration?}, \apj~ {\bf 649} (2006) 563, [\texttt
{arXiv:astro-ph/0512586}].

\bibitem{blandford} Blandford R. et al., {\it Cosmokinetics}, {\it
Observing Dark Energy (NOAO/Tucson proceedings)}, [\texttt
{arXiv:astro-ph/0408279}].

\bibitem{turner} Turner M. \& Riess  A., {\it Do SNe Ia Provide Direct Evidence for
Past Deceleration of the Universe?}, \apj~ {\bf 569} (2002) 18,
[\texttt {arXiv:astro-ph/0106051}].

\bibitem{dodelson} Dodelson S. et al., {\it Solving the Coincidence Problem: Tracking
Oscillating Energy}, \prl~ {\bf 85} (2000) 5276, [\texttt
{arXiv:astro-ph/0002360}].

\bibitem{german} German G. \& de la Macorra A., {\it Lect. Notes Phys}~ {\bf 646}
(2004) 259.

\bibitem{re} Riess A. G. et al., (Supernova Search Team), {\it Type Ia Supernova
Discoveries at $z>1$ From the Hubble Space Telescope: Evidence for
Past Deceleration and Constraints on Dark Energy Evolution}, \apj~
{\bf 607} (2004) 665, [\texttt {arXiv:astro-ph/0402512}].

\bibitem{komatsu09} Komatsu E. et al., {\it Five-Year Wilkinson
Microwave Anisotropy Probe Observations: Cosmological
Interpretation}, \apjs~ {\bf 180} (2009) 330, [\texttt
{arXiv:0803.0547}].

\bibitem{tarundeep} Saini T. D. et al., {\it Reconstructing the
cosmic equation of state from supernova distances},  \prl~ {\bf 85}
(2000) 1162, [\texttt {arXiv:astro-ph/9910231}].

\bibitem{sim} Simon J., Verde L. \& Jimenez R.,
{\it Constraints on the redshift dependence of the dark energy
potential}, \prd \ {\bf 71} (2005) 123001, [\texttt
{arXiv:astro-ph/0412269}].

\bibitem{dan} Dantas M. A., Alcaniz J. S. Jain D. \& Dev A.,
{\it Age constraints on the cosmic equation of state}, \AnA~  {\bf
467} (2007) 421, [\texttt {arXiv:astro-ph/0607060}].

\bibitem{st} Stern D. et al., {\it Cosmic chronometers: constraining the equation
of state of dark energy. I: H(z) measurements}, JCAP {\bf 1002}
(2010) 008, [\texttt {arXiv:0907.3152}].

\bibitem{gz} Gaztanaga E., Cabr A. \& Hui A., {\it Clustering of luminous red galaxies - IV.
Baryon acoustic peak in the line-of-sight direction and a direct
measurement of H(z)}, \mnras \ {\bf 399} (2009) 1663, [\texttt
{arXiv:0807.3551}].

\bibitem{rat} Chen Y.\& Ratra B., {\it Hubble parameter data constraints on dark energy},
[\texttt {arXiv:1106.4294}].

\bibitem{zh} Zhang T. J. et al., {\it Constraints on the Dark Side of
the Universe and Observational Hubble Parameter Data},
 Adv. Astron~ (2010) 184284, [\texttt {arXiv:1010.1307}].

\bibitem{ma} Ma C. \& Zhang T. J., {\it Power of Observational Hubble
Parameter Data: A Figure of Merit Exploration}, \apj~ {\bf 730}
(2011) 74, [\texttt {arXiv:1007.3787}].

\bibitem{aman} Amanullah R. et al., {\it Spectra and Light Curves of Six Type
Ia Supernovae at 0.511$< z< $1.12 and the Union2 Compilation}, \apj~
{\bf 716} (2010) 712, [\texttt {arXiv:1004.1711}].

\bibitem{Cooray} Cooray A. et al., {\it Measuring Angular Diameter Distances
through Halo Clustering} \apjl~ {\bf 557} (2001) 7, [\texttt
{arXiv:astro-ph/0105061}].

\bibitem{wz1} Blake C. et al., {\it The WiggleZ Dark Energy Survey: mapping the
distance-redshift relation with baryon acoustic oscillations},
[\texttt {arXiv:1108.2635}].

\bibitem{wz2} Blake C. et al., {\it The WiggleZ Dark Energy Survey: testing
the cosmological model with baryon acoustic oscillations at z=0.6},
[\texttt {arXiv:1105.2862}].

\bibitem{percival} Percival W. J. et al., {\it Baryon acoustic
oscillations in the Sloan Digital Sky Survey Data Release 7 galaxy
sample}, \mnras~  {\bf 401} (2010) 2148, [\texttt
{arXiv:0907.1660}].

\bibitem{beutler} Beutler F. et al., {\it The 6dF Galaxy Survey:
baryon acoustic oscillations and the local Hubble constant}, \mnras~
{\bf 416} (2011) 3017, [\texttt {arXiv:1106.3366}].

\bibitem{komatsu11}  Komatsu E. et al., {\it Seven-year Wilkinson Microwave Anisotropy
Probe (WMAP) Observations: Cosmological Interpretation},  \apjs~
{\bf 192} (2011)  18, [\texttt {arXiv:1001.4538}].

\bibitem{sivia} Sivia D. S., {\it Data analysis: A Bayesian tutorial, Oxford
Science Publication}.

\bibitem{elgaroy} Elgaroy O. \& Multamaki T., {\it Bayesian analysis of Friedmannless
cosmologies}, JCAP {\bf 9} (2006) 2, [\texttt
{arXiv:astro-ph/0603053}].

\bibitem{gong} Gong Y. \& Wang A., {\it Observational constraints on the acceleration
of the Universe}, \prd~ {\bf 73} (2006) 083506, [\texttt
{arXiv:astro-ph/0601453}].

\bibitem{cun} Cunha J. V. \& Lima J. A. S., {\it Transition redshift: new
kinematic constraints from supernovae}, \mnras~ {\bf 390} (2008)
210, [\texttt {arXiv:0805.1261}].

\bibitem{gui} Guimaraes A.C.C., Cunha J.V. \& Lima J.A.S., {\it Bayesian Analysis
and Constraints on Kinematic Models from Union SNIa}, JCAP {\bf 10}
(2009) 010, [\texttt {arXiv:0904.3550}].

\bibitem{rapetti} Rapetti D. et al., {\it A kinematical approach to dark energy
studies}, \mnras~ {\bf 375} (2007) 1510, [\texttt
{arXiv:astro-ph/0605683}].

\bibitem{lim} Holanda R.F.L. et al., {\it Accessing the Acceleration
of the Universe with Sunyaev-Zel'dovich and X-ray Data from Galaxy
Clusters}, [\texttt {arXiv:1103.2688}].

\bibitem{mort} Mortsell E. \& Clarkson C., {\it Model independent constraints on the
cosmological expansion rate}, JCAP {\bf 01} (2009) 44, [\texttt
{arXiv:0811.0981}]

\bibitem{xu} Xu L. et al., {\it Constraints on Kinematic Model from Recent
Cosmic Observations: SNe Ia, BAO and Observational Hubble Data},
JCAP {\bf 07} (2009) 031, [\texttt {arXiv:0905.4552}].

\bibitem{lu} Lu J. et al., {\it Constraints on kinematic models from the
 latest observational data}, \plb ~ {\bf 699} (2011) 246,
 [\texttt {arXiv:1105.1871}].



\end{thebibliography}
\end{document}